\documentclass[aps,pra,twocolumn,showpacs,floatfix]{revtex4-1}

\usepackage{graphicx}
\usepackage{color}

\begin{document}

\title{Creating Feshbach resonances for ultracold molecule formation \\ with radiofrequency fields}

\author{Daniel J. Owens}
\author{Ting Xie}
\author{Jeremy M. Hutson}
\email{Author to whom correspondence should be addressed; j.m.hutson@durham.ac.uk}
\affiliation{Joint Quantum Centre (JQC) Durham-Newcastle, Department of
Chemistry, Durham University, South Road, Durham DH1 3LE, United Kingdom.}
\date{\today}

\begin{abstract}
We show that radiofrequency (RF) radiation may be used to create Feshbach
resonances in ultracold gases of alkali-metal atoms at desired magnetic fields
that are convenient for atomic cooling and degeneracy. For the case of
$^{39}$K+$^{133}$Cs, where there are no RF-free resonances in regions where Cs
may be cooled to degeneracy, we show that a resonance may be created near 21~G
with 69.2~MHz RF radiation. This resonance is almost lossless with circularly
polarized RF, and the molecules created are long-lived even with
plane-polarized RF.
\end{abstract}
\date{\today}
%\pacs{37.10.Mn, 34.50.Cx}
%34.50.Cx	Elastic; ultracold collisions
%37.10.Mn	Slowing and cooling of molecules
\maketitle

\section{Introduction}

Polar molecules formed from ultracold atoms are opening up new possibilities
for quantum-controlled chemistry \cite{Krems:PCCP:2008}, precision measurement
\cite{Flambaum:2007, Hudson:2011, Baron:2014}, quantum computation
\cite{DeMille:2002}, quantum phase transitions \cite{Wall:2010} and quantum
simulation \cite{Micheli:2006, Baranov:2012}. The last few years have seen
major success, with the formation of ultracold $^{40}$K$^{87}$Rb
\cite{Ni:KRb:2008}, $^{87}$Rb$^{133}$Cs \cite{Takekoshi:RbCs:2014,
Molony:RbCs:2014}, $^{23}$Na$^{40}$K \cite{Park:2015} and most recently
$^{23}$Na$^{87}$Rb \cite{Guo:NaRb:2016} molecules in their absolute ground
states. Molecules are first formed by magnetoassociation, in which atom pairs
are converted into weakly bound molecules by ramping a magnetic field across a
magnetically tunable Feshbach resonance. The resulting ``Feshbach molecules"
are then transferred to the polar ground state by stimulated Raman adiabatic
passage (STIRAP). The ground-state molecules have been confined in
one-dimensional \cite{deMiranda:2011} and three-dimensional \cite{Chotia:2012}
optical lattices and used to study atom-molecule and molecule-molecule
collision processes \cite{Ospelkaus:react:2010, Takekoshi:RbCs:2014}.

A major problem in this field is that the magnetoassociation step is possible
only if there is a Feshbach resonance of suitable width at a magnetic field
where there is a lucky combination of intraspecies and interspecies scattering
lengths. Ideally, all three scattering lengths have moderate positive values to
allow cooling, condensate formation and mixing of the two atomic clouds. For
the intraspecies scattering lengths, negative values cause condensate collapse,
whereas excessively positive values cause loss through fast 3-body
recombination. For the interspecies scattering length, a large negative value
can cause collapse of the mixed condensate, while a large positive value can
make the condensates of the two species immiscible. Although magnetoassociation
can be carried out in low-temperature thermal gases that are not subject to
condensate collapse, it is much less efficient than in condensates and does not
produce high densities of molecules. This is the so-called {\em one-field
problem}, because a single field must be chosen to satisfy several different
criteria, and such a field may not (often does not) exist.

The purpose of this Letter is to show that radiofrequency (RF) fields can be
used to produce new Feshbach resonances that offer additional possibilities for
magnetoassociation. In particular, they may be used to produce resonances at
magnetic fields where the scattering lengths have desired properties. Formally
similar resonances have been considered previously in homonuclear systems
\cite{TVTscherbul:rf:2010, Hanna:2010, Smith:rf:2015}, and molecules have been
formed by direct RF association \cite{Thompson:magres:2005, Zirbel:2008}. We
propose here that rf-induced resonances may provide a solution to the one-field
problem in heteronuclear systems.

We recently considered the possibilities for magnetoassociation to form
molecules in mixtures of $^{39}$K, $^{40}$K and $^{41}$K with $^{133}$Cs
\cite{Patel:2014} by performing coupled-channel calculations of the Feshbach
resonance positions and widths, using interaction potentials obtained from
extensive spectroscopic studies \cite{Ferber:2013}. In all three systems, we
found Feshbach resonances with widths suitable for magnetoassociation. However,
the background intraspecies and interspecies scattering lengths around the
resonances present problems. In particular, the intraspecies scattering length
for $^{133}$Cs is very large and positive except in relatively narrow windows
around 21~G, 559~G and 894~G \cite{Berninger:Cs2:2013}, and for $^{39}$KCs and
$^{40}$KCs there were no suitable interspecies Feshbach resonances that lie in
these regions. In the present work, we show for the case of $^{39}$KCs that a
suitable RF field can be used to create a new Feshbach resonance in the
magnetic field region near 21~G, where Cs can be cooled to condensation.

\begin{figure*}[t]
\includegraphics[width=\textwidth]{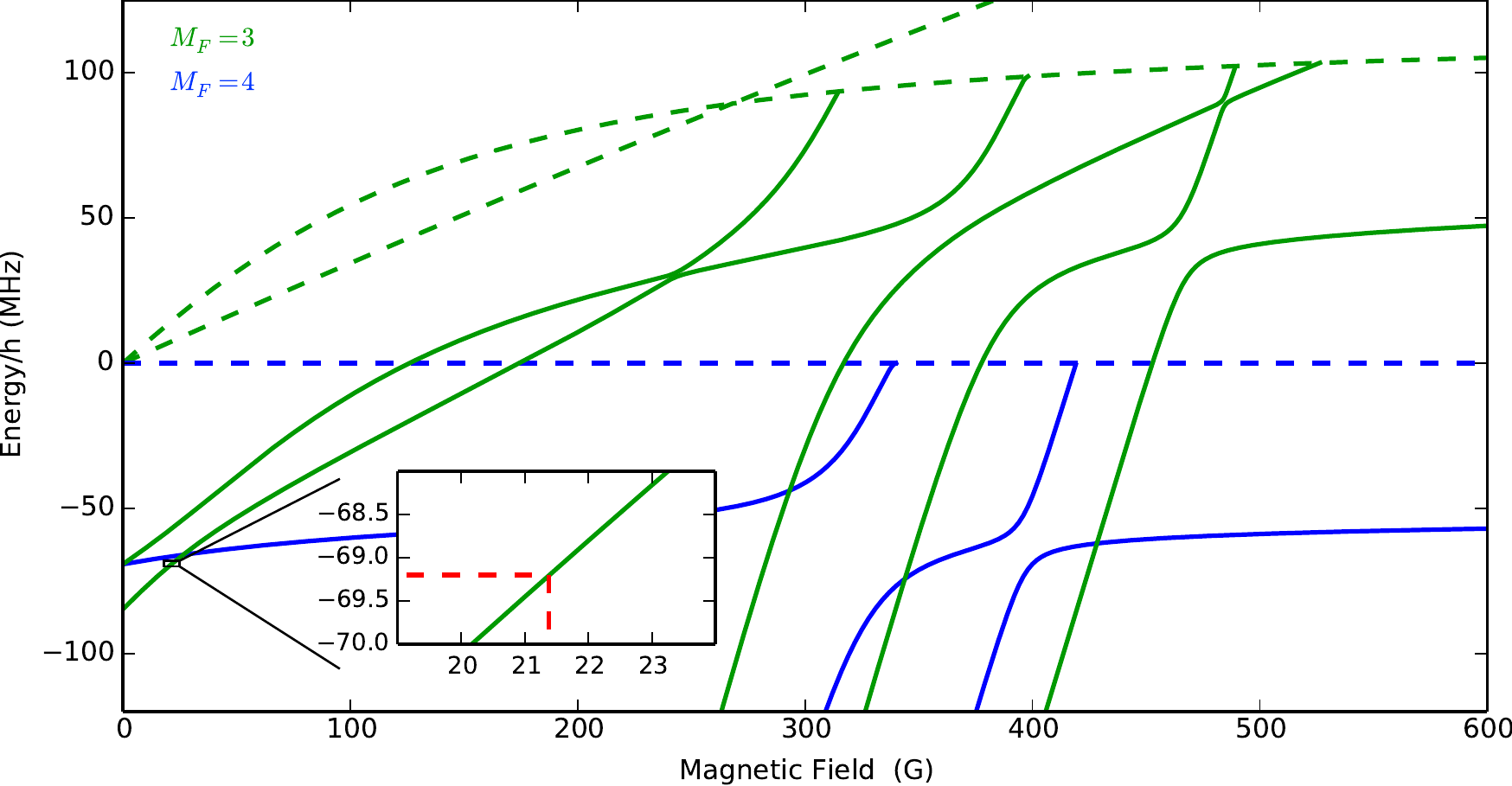}
\caption{(Color online) Thresholds (dashed lines) and near-threshold bound
states (solid lines) for $^{39}$KCs in the absence of RF radiation for $M_F=+4$
(blue) and $M_F=+3$ (orange). The inset shows an expanded view of the region we
consider in detail. All energies are relative to the lowest $M_F=+4$
threshold.} \label{fig:bound}
\end{figure*}

\section{Methods}

For the present work, we have generalized the MOLSCAT \cite{molscat:v14}, BOUND
\cite{Hutson:bound:1993} and FIELD \cite{Hutson:field:2011} programs to handle
interactions of two alkali-metal atoms in the presence of simultaneous magnetic
and RF fields. MOLSCAT performs scattering calculations to extract S matrices
and scattering lengths, and locate and characterize Feshbach resonances. BOUND
locates near-threshold bound states as a function of energy at constant applied
magnetic and RF field. The extended version of FIELD is capable of locating
bound states at fixed energy as a function of magnetic field, RF field
strength, or RF frequency. Both scattering and bound-state calculations use
propagation methods that do not rely on basis sets in the interatomic distance
coordinate $R$. Apart from the inclusion of RF fields, which is new in the
present work, the coupled-channel methodology is the same as described for Cs
in Section IV of Ref.\ \cite{Berninger:Cs2:2013}, so only a brief summary will
be given here.

We use a basis set of photon-dressed products of atomic functions in a fully
decoupled representation, $|s_a m_{sa}\rangle |i_a m_{ia}\rangle |s_b
m_{sb}\rangle |i_b m_{ib}\rangle |L M_L\rangle |N M_N\rangle$, where $s_a$ and
$s_b$ are the electron spins of the two atoms, $i_a$ and $i_b$ are their
nuclear spins, $L$ is the angular momentum of their relative motion, and $N$ is
the photon number with respect to the average photon number $N_0$. The
quantities $m$ and $M$ are the corresponding projections onto the magnetic
field axis $Z$. The Hamiltonian and its matrix elements in this basis set have
been given in the Appendix of ref.~\cite{Hutson:Cs2-note:2008}, except for the
RF terms, which are described below.

The calculation may be done for a variety of different polarizations of the RF
radiation. For radiation polarized along $Z$ ($\pi$ polarization), $M_N=0$ for
all $N$ and $M_{\cal F}=M_F+M_L$ is conserved, where
$M_F=m_{sa}+m_{ia}+m_{sb}+m_{ib}$. For radiation polarized in the XY plane, the
simplest calculation is for circularly polarized light, with either $M_N=N$
(right-circularly polarized, $\sigma_+$) or $M_N=-N$ (left-circularly
polarized, $\sigma_-$). For radiation linearly polarized along $X$
($\sigma_X$), $M_N$ runs from $-N$ to $N$ in steps of 2 and a correspondingly
larger basis set is required. In all these cases, $M_{\rm tot}=M_{\cal F}+M_N$
is conserved. In the present work we restrict the basis set to functions with
$|N|\le 2$ and the required $M_{\rm tot}$.

The RF terms in the Hamiltonian for each atom are given for $\sigma_+$
polarization by
\begin{eqnarray}
H_{\rm rf}&=&\frac{\mu_{\rm B} B_{\rm rf}}{2\sqrt{N}} \left[(g_S\hat s_+ +
g_i\hat i_+)\hat a_+ + (g_S\hat s_- + g_i\hat i_-)\hat a_+^\dagger\right]\cr
&+&h\nu(\hat{a}_+\hat{a}_+^\dagger-N_0)
\end{eqnarray}
where $B_{\rm rf}$ is the oscillating magnetic field, $\nu$ is the RF
frequency, $\hat s_+$ and $\hat s_-$ are raising and lowering operators for the
electron spin, $\hat i_+$ and $\hat i_-$ are the corresponding operators for
the nuclear spin, and $g_S$ and $g_i$ are electron and nuclear spin $g$-factors
with the sign convention of Arimondo {\em et al.} \cite{Arimondo:1977}.
$\hat{a}_+$ and $\hat{a}_+^\dagger$ are photon annihilation and creation
operators for $\sigma_+$ photons. For $\sigma_-$ polarization, $\hat a_-$
replaces $\hat a_+^\dagger$ and $\hat a_-^\dagger$ replaces $\hat a_+$. For
$\sigma_X$ polarization, both $\sigma_+$ and $\sigma_-$ coupling terms are
present, renormalized by $1/\sqrt{2}$.

\section{Results}

Figure \ref{fig:bound} shows the near-threshold $L=0$ bound states of
$^{39}$KCs, in the absence of RF radiation, for both $M_F=+4$, corresponding to
$^{39}$K and $^{133}$Cs atoms in their absolute ground states, and $M_F=+3$.
All levels are shown relative to the lowest $M_F=+4$ threshold, and the two
$M_F=+3$ thresholds are shown as dotted orange lines. At fields near 21~G,
where the scattering length of Cs allows cooling to condensation, it may be
seen that there are $M_F=+3$ bound states that lie about $-57$ and $-69$ MHz
below the $M_F=+4$ threshold.

\begin{figure}[t]
\includegraphics[width=\columnwidth]{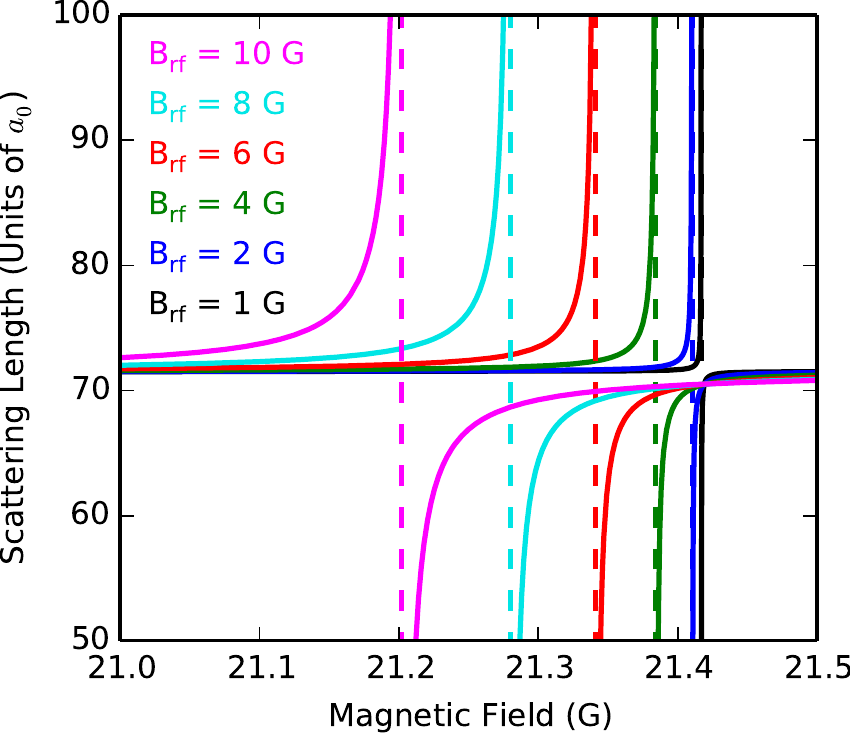}
\caption{(Color online) Calculated scattering length for $^{39}$K+$^{133}$Cs,
in the presence of a $\sigma_+$ RF field at a frequency of 69.2~MHz, with
differing strengths $B_{\rm rf}$ (increasing from right to
left).}\label{fig:res}
\end{figure}

We choose an RF frequency of 69.2~MHz to bring one of the $M_F=+3$ states into
resonance with the $M_F=+4$ threshold near 21~G and carry out scattering
calculations in the field-dressed basis set for $M_{\rm tot} = +4$ to identify
Feshbach resonances. Fig.\ \ref{fig:res} shows the calculated interspecies
scattering length for $^{39}$K+$^{133}$Cs collisions in the region around 21~G
for a variety of strengths $B_{\rm rf}$ of the RF field, with $\sigma_+$
polarization and $L_{\max}=0$. It may be seen that a new resonance is induced,
with a width that varies approximately quadratically with RF field. To a good
approximation the width $\Delta$ is $1.6\times10^{-5}$ $B_{\rm rf}^2$/G. The
RF-induced resonance is also shifted significantly from its RF-free position,
again nearly quadratically with field.

The RF fields considered in this paper are large, but comparable to those
considered previously \cite{TVTscherbul:rf:2010, Hanna:2010}. RF fields up to
6~G have been applied in experiments to produce $^{87}$Rb$_2$ on atom chips,
and higher fields are achievable \cite{Mordovin:2015}. The fields currently
achievable in conventional atom traps are rather lower, but fields of up to 0.7
G have been achieved \cite{Morizot:2008}.

The resonances shown in Fig.\ \ref{fig:res} are lossless, so appear as true
poles in the scattering length. This is because the incoming channel is the
lowest that exists for $M_{\rm tot}=4$ and the molecular state that is coupled
to it by RF radiation is a true bound state, below the lowest threshold.
However, there are two decay mechanisms that can actually exist. First, if the
RF radiation has $\sigma_X$ rather than $\sigma_+$ polarization, it can couple
to an $M_{\rm tot}=4$ channel with $M_F=3, L=0, N=-1, M_N=+1$. Because $N=-1$,
this lies below the incoming channel. The resonance is then characterized by a
resonant scattering length $a_{\rm res}$ in addition to the width $\Delta$: the
real part of the scattering length exhibits an oscillation of amplitude $\pm
a_{\rm res}/2$ instead of a pole, and the imaginary part exhibits a narrow peak
of magnitude $\pm a_{\rm res}$ \cite{Hutson:res:2007}. We have repeated the
calculations of Fig.\ \ref{fig:res} for $\sigma_X$ polarization, and find
$a_{\rm res}=1.5\times10^7 ({\rm G}/B_{\rm rf})^2$ bohr. These very large
values of $a_{\rm res}$ correspond to very weakly decayed resonances, and
should not cause problems in magnetoassociation. Secondly, even for $\sigma_+$
polarization, channels with $L>0$ and $M_L\ne0$ can cause collisionally
assisted one-photon decay, mediated by the atomic spin dipolar (or second-order
spin-orbit) interaction. In the present case, for example, there is a channel
$M_F=3, L=2, M_L=+2, N=-1, M_N=-1$, and thus $M_{\cal F}=5, M_{\rm tot}=4$,
that lies below the incoming channel. Such d-wave participation can in
principle cause loss. However, this is a very weak process because of the
weakness of the spin-dipolar coupling. We have repeated the calculations of
Fig.\ \ref{fig:res} with all $L=2$ channels for $M_{\rm tot}=4$ included; in
this case the resonance is close to pole-like with $a_{\rm res}=1.2\times10^8$
bohr for $B_{\rm rf}=10$~G. Once again, therefore, this loss process should not
cause problems in magnetoassociation.

The resonant scattering length $a_{\rm res}$ is given by \cite{Hutson:res:2007}
\begin{equation}
a_{\rm res}=-2a_{\rm bg}\Delta/\Gamma^B_{\rm inel},
\end{equation}
where $\Gamma^B_{\rm inel}$ is a Breit-Wigner width that describes decay of the
field-dressed bound state to atoms. This may be converted into a lifetime for
the field-dressed molecules,
\begin{equation}
\tau=\left|\frac{\hbar}{\Gamma^B_{\rm inel}\Delta\mu}\right|
= \left|\frac{-\hbar a_{\rm res}}{2\Delta\mu a_{\rm bg}\Delta}\right|,
\end{equation}
where $\Delta\mu$ is the difference in magnetic moments between the molecular
state and the incoming channel, $\Delta\mu=\mu_{\rm molecule}-\mu_{\rm atoms}$.
The value $a_{\rm res}=1.5\times10^5$ bohr obtained for $\sigma_X$ polarization with
$B_{\rm rf}=10$~G corresponds to a molecular lifetime of 166~ms for
photon-assisted decay to the lower field-dressed threshold; the lifetime is
approximately proportional to $B_{\rm rf}^{-4}$, as expected for a 2-photon
decay pathway, so increases fast as the RF field is decreased. This decay of
course persists only as long as the RF field is switched on.

Different type of decaying RF-induced resonance may be observed if the RF
radiation couples the incoming state to a molecular state that is itself above
a threshold to which it can decay. At least two such cases may be identified.
Tscherbul \emph{et al.}\ \cite{TVTscherbul:rf:2010} and Hanna \emph{et al.}\
\cite{Hanna:2010} both considered RF-induced resonances due to bound states of
$^{87}$Rb$_2$ near the a+e $|1,1\rangle+|2,-1\rangle$ excited hyperfine
threshold of $^{87}$Rb; these bound states can decay to lower open channels
with the same $M_F$ through RF-independent mechanisms, so the resonances are
strongly decayed and the molecules have a finite lifetime even after the RF
field is switched off. Hanna \emph{et al.}\ \cite{Hanna:2010} also considered
resonances due to bound states of $^6$Li$_2$ that lie above the lowest open
channel, but have different $M_F$; these can decay to $L=2$ open channels by
RF-free spin-dipolar coupling, or through 2-photon RF coupling for $\sigma_X$
polarization.

The coupled-channel approach that we use includes the effect of the RF field
nonperturbatively. However, for the RF fields considered here, the resonance
widths are clearly dominated by direct couplings from the incoming channel to
the resonant bound state. Under these circumstances, the width of the resonance
is proportional to the square of a bound-continuum matrix element $I$ of the RF
perturbation $\hat H_{\rm RF}$,
\begin{equation}
\Delta = \frac{\pi I^2}{k\Delta\mu a_{\rm bg}},
\end{equation}
where
\begin{equation}
I= \left\langle \psi_{\rm bound} \right| \hat H_{\rm RF}
\left| \psi_{\rm incoming} \right\rangle.
\end{equation}
The incoming wave function is essentially a product of field-dressed atomic
functions $|\alpha_{\rm K}m_{f,{\rm K}}\rangle$ and $|\alpha_{\rm Cs}m_{f,{\rm
Cs}}\rangle$ and a radial function $\chi_k(r)$. At the low magnetic fields
considered here, the atomic functions are approximately
$|f,m_f\rangle=|1,1\rangle$ for $^{39}$K and $|3,3\rangle$ for $^{133}$Cs,
where $f$ is the resultant of $s$ and $i$ for each atom. The molecular wave
functions are more complicated, and a general treatment is beyond the scope of
this letter. However, for the specific case of $^{39}$K$^{133}$Cs, Fig.\
\ref{fig:bound} shows that the near-threshold bound states are mostly nearly
parallel to the thresholds, indicating that they have similar spin character to
the thresholds where this is true. If the scattering lengths for the $M_F=+3$
and $+4$ thresholds were identical, the incoming and bound-state radial
functions would be orthogonal to one another, which would produce a very small
matrix element $I$ because the spin part of the RF coupling is almost
independent of $r$. In general terms, therefore, the RF coupling is strongest
for systems where the scattering lengths for the incoming and bound-state
channels differ most, and thus where the singlet and triplet scattering lengths
are very different. It is reasonably straightforward to construct a complete
map of the near-threshold bound states for any specific system using BOUND and
FIELD, but some experimentation is needed to establish which bound states
produce RF-induced resonances with useful widths.

\begin{figure}[t]
\includegraphics[width=\columnwidth]{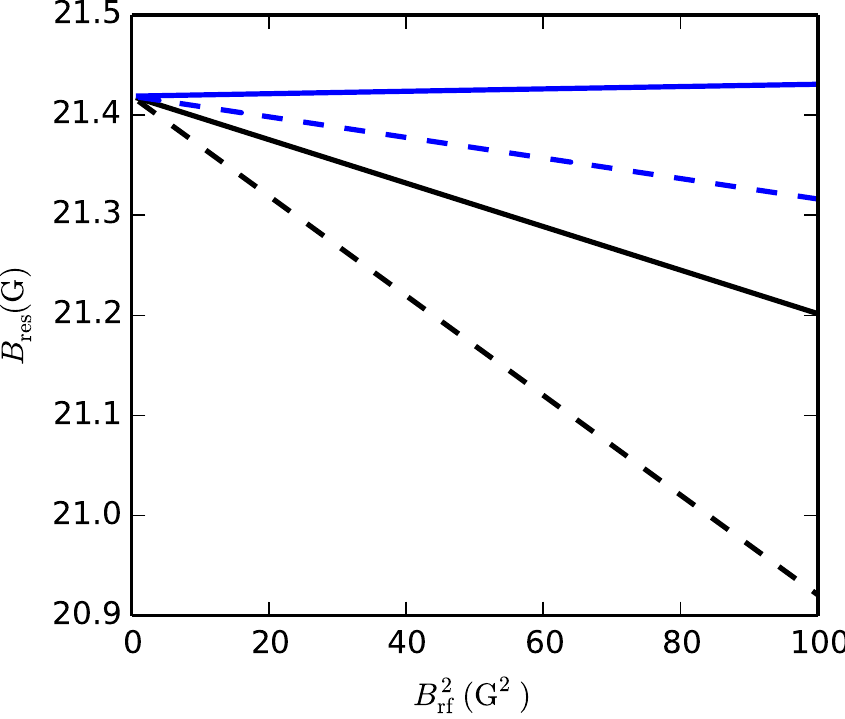}
\caption{(Color online) Calculated resonance positions as a function of $B_{\rm
rf}^2$ for $\sigma_+$ (black lines) and $\sigma_X$ polarization
(orange lines), for basis sets with $|N|\le1$ (dashed lines) and $|N|\le2$ (solid
lines).} \label{fig:shifts}
\end{figure}

Although the resonance widths are dominated by direct couplings between the
incoming channel and the resonant bound state, the shifts are not. Figure
\ref{fig:shifts} shows the resonance positions as a function of $B_{\rm rf}^2$
for both $\sigma_+$ and $\sigma_X$ polarization, for basis sets with both
$|N|\le2$ (essentially converged) and $|N|\le1$ (unconverged). The smaller
basis sets give widths that are unchanged to 1 part in $10^3$ compared to the
larger ones, but the resonance positions shift substantially; they are still
close to quadratic in $B_{\rm rf}$, but with different coefficients. This
arises because the $M_F=+3$, $N=1$ bound state that causes the resonances is
shifted by ac-Zeeman couplings to both $N=0$ and $N=2$ states, but the latter
couplings are omitted for the smaller basis sets. The shifts are also
significantly different for the two polarizations. Our coupled-channel approach
provides a straightforward way to capture such effects properly.

Resonances of the type described here will exist for all the alkali-metal
dimers. For all such dimers except those containing $^{40}$K, the lowest
threshold in a magnetic field has $M_{F,{\rm ground}}=i_a+i_b-1$. For those
containing $^{40}$K, which has inverted hyperfine structure, $M_{F,{\rm
ground}}=i_a+i_b$. In both cases, there are Zeeman-excited thresholds with
$M_F<M_{F,{\rm ground}}$. However, the lowest thresholds with $M_F>M_{F,{\rm
ground}}$ always correlate with excited hyperfine states and are substantially
higher in energy. As for $^{39}$K$^{133}$Cs, resonances due to bound states
with $M_F=M_{F,{\rm ground}}-1$ are likely to be pole-like, with only weak
decay as described above.

\section{Conclusions}

We have shown that radiofrequency fields can be used to engineer magnetically
tunable Feshbach resonances in regions of magnetic field where they did not
previously exist. This capability may allow the creation of resonances at
magnetic fields where the intraspecies and interspecies scattering lengths have
values that are favorable for evaporative or sympathetic cooling, and where
stable mixed condensates may be created. This in turn may allow
magnetoassociation to form molecules from otherwise intractable pairs of
ultracold atoms. The resonances we consider are different from those of refs.\
\cite{TVTscherbul:rf:2010} and \cite{Hanna:2010}, both because the molecules
that can be created at them are heteronuclear and because they are truly bound,
so cannot decay to lower atomic thresholds after the RF radiation is switched
off.

The present work has used an RF field to bring bring bound states into
resonance with threshold and create new Feshbach resonances. This is
conceptually the simplest approach, but a similar effect could be achieved with
the difference between two laser frequencies, with different (and potentially
more versatile) selection rules governing which bound states can cause
resonances.
%More sophisticated combinations of applied fields can also be envisaged.

\acknowledgments

The authors are grateful to Caroline Blackley for initial work on adapting the
computer programs to handle RF fields. This work has been supported by the UK
Engineering and Physical Sciences Research Council (Grants No. ER/I012044/1 and
EP/N007085/1 and a Doctoral Training Partnership with Durham University).

\bibliography{../all}

\begin{thebibliography}{32}%
\makeatletter
\providecommand \@ifxundefined [1]{%
 \@ifx{#1\undefined}
}%
\providecommand \@ifnum [1]{%
 \ifnum #1\expandafter \@firstoftwo
 \else \expandafter \@secondoftwo
 \fi
}%
\providecommand \@ifx [1]{%
 \ifx #1\expandafter \@firstoftwo
 \else \expandafter \@secondoftwo
 \fi
}%
\providecommand \natexlab [1]{#1}%
\providecommand \enquote  [1]{``#1''}%
\providecommand \bibnamefont  [1]{#1}%
\providecommand \bibfnamefont [1]{#1}%
\providecommand \citenamefont [1]{#1}%
\providecommand \href@noop [0]{\@secondoftwo}%
\providecommand \href [0]{\begingroup \@sanitize@url \@href}%
\providecommand \@href[1]{\@@startlink{#1}\@@href}%
\providecommand \@@href[1]{\endgroup#1\@@endlink}%
\providecommand \@sanitize@url [0]{\catcode `\\12\catcode `\$12\catcode
  `\&12\catcode `\#12\catcode `\^12\catcode `\_12\catcode `\%12\relax}%
\providecommand \@@startlink[1]{}%
\providecommand \@@endlink[0]{}%
\providecommand \url  [0]{\begingroup\@sanitize@url \@url }%
\providecommand \@url [1]{\endgroup\@href {#1}{\urlprefix }}%
\providecommand \urlprefix  [0]{URL }%
\providecommand \Eprint [0]{\href }%
\providecommand \doibase [0]{http://dx.doi.org/}%
\providecommand \selectlanguage [0]{\@gobble}%
\providecommand \bibinfo  [0]{\@secondoftwo}%
\providecommand \bibfield  [0]{\@secondoftwo}%
\providecommand \translation [1]{[#1]}%
\providecommand \BibitemOpen [0]{}%
\providecommand \bibitemStop [0]{}%
\providecommand \bibitemNoStop [0]{.\EOS\space}%
\providecommand \EOS [0]{\spacefactor3000\relax}%
\providecommand \BibitemShut  [1]{\csname bibitem#1\endcsname}%
\let\auto@bib@innerbib\@empty
%</preamble>
\bibitem [{\citenamefont {Krems}(2008)}]{Krems:PCCP:2008}%
  \BibitemOpen
  \bibfield  {author} {\bibinfo {author} {\bibfnamefont {R.~V.}\ \bibnamefont
  {Krems}},\ }\href@noop {} {\bibfield  {journal} {\bibinfo  {journal} {Phys.
  Chem. Chem. Phys.}\ }\textbf {\bibinfo {volume} {10}},\ \bibinfo {pages}
  {4079} (\bibinfo {year} {2008})}\BibitemShut {NoStop}%
\bibitem [{\citenamefont {Flambaum}\ and\ \citenamefont
  {Kozlov}(2007)}]{Flambaum:2007}%
  \BibitemOpen
  \bibfield  {author} {\bibinfo {author} {\bibfnamefont {V.~V.}\ \bibnamefont
  {Flambaum}}\ and\ \bibinfo {author} {\bibfnamefont {M.~G.}\ \bibnamefont
  {Kozlov}},\ }\href@noop {} {\bibfield  {journal} {\bibinfo  {journal} {Phys.
  Rev. Lett.}\ }\textbf {\bibinfo {volume} {99}},\ \bibinfo {pages} {150801}
  (\bibinfo {year} {2007})}\BibitemShut {NoStop}%
\bibitem [{\citenamefont {Hudson}\ \emph {et~al.}(2011)\citenamefont {Hudson},
  \citenamefont {Kara}, \citenamefont {Smallman}, \citenamefont {Sauer},
  \citenamefont {Tarbutt},\ and\ \citenamefont {Hinds}}]{Hudson:2011}%
  \BibitemOpen
  \bibfield  {author} {\bibinfo {author} {\bibfnamefont {J.~J.}\ \bibnamefont
  {Hudson}}, \bibinfo {author} {\bibfnamefont {D.~M.}\ \bibnamefont {Kara}},
  \bibinfo {author} {\bibfnamefont {J.}~\bibnamefont {Smallman}}, \bibinfo
  {author} {\bibfnamefont {B.~E.}\ \bibnamefont {Sauer}}, \bibinfo {author}
  {\bibfnamefont {M.~R.}\ \bibnamefont {Tarbutt}}, \ and\ \bibinfo {author}
  {\bibfnamefont {E.~A.}\ \bibnamefont {Hinds}},\ }\href@noop {} {\bibfield
  {journal} {\bibinfo  {journal} {Nature London}\ }\textbf {\bibinfo {volume}
  {473}},\ \bibinfo {pages} {493} (\bibinfo {year} {2011})}\BibitemShut
  {NoStop}%
\bibitem [{\citenamefont {Baron}\ \emph {et~al.}(2014)\citenamefont {Baron},
  \citenamefont {Campbell}, \citenamefont {DeMille}, \citenamefont {Doyle},
  \citenamefont {Gabrielse}, \citenamefont {Gurevich}, \citenamefont {Hess},
  \citenamefont {Hutzler}, \citenamefont {Kirilov}, \citenamefont {Kozyryev},
  \citenamefont {O'Leary}, \citenamefont {Panda}, \citenamefont {Parsons},
  \citenamefont {Petrik}, \citenamefont {Spaun}, \citenamefont {Vutha},\ and\
  \citenamefont {West}}]{Baron:2014}%
  \BibitemOpen
  \bibfield  {author} {\bibinfo {author} {\bibfnamefont {J.}~\bibnamefont
  {Baron}}, \bibinfo {author} {\bibfnamefont {W.~C.}\ \bibnamefont {Campbell}},
  \bibinfo {author} {\bibfnamefont {D.}~\bibnamefont {DeMille}}, \bibinfo
  {author} {\bibfnamefont {J.~M.}\ \bibnamefont {Doyle}}, \bibinfo {author}
  {\bibfnamefont {G.}~\bibnamefont {Gabrielse}}, \bibinfo {author}
  {\bibfnamefont {Y.~V.}\ \bibnamefont {Gurevich}}, \bibinfo {author}
  {\bibfnamefont {P.~W.}\ \bibnamefont {Hess}}, \bibinfo {author}
  {\bibfnamefont {N.~R.}\ \bibnamefont {Hutzler}}, \bibinfo {author}
  {\bibfnamefont {E.}~\bibnamefont {Kirilov}}, \bibinfo {author} {\bibfnamefont
  {I.}~\bibnamefont {Kozyryev}}, \bibinfo {author} {\bibfnamefont {B.~R.}\
  \bibnamefont {O'Leary}}, \bibinfo {author} {\bibfnamefont {C.~D.}\
  \bibnamefont {Panda}}, \bibinfo {author} {\bibfnamefont {M.~F.}\ \bibnamefont
  {Parsons}}, \bibinfo {author} {\bibfnamefont {E.~S.}\ \bibnamefont {Petrik}},
  \bibinfo {author} {\bibfnamefont {B.}~\bibnamefont {Spaun}}, \bibinfo
  {author} {\bibfnamefont {A.~C.}\ \bibnamefont {Vutha}}, \ and\ \bibinfo
  {author} {\bibfnamefont {A.~D.}\ \bibnamefont {West}},\ }\href@noop {}
  {\bibfield  {journal} {\bibinfo  {journal} {Science}\ }\textbf {\bibinfo
  {volume} {343}},\ \bibinfo {pages} {269} (\bibinfo {year}
  {2014})}\BibitemShut {NoStop}%
\bibitem [{\citenamefont {DeMille}(2002)}]{DeMille:2002}%
  \BibitemOpen
  \bibfield  {author} {\bibinfo {author} {\bibfnamefont {D.}~\bibnamefont
  {DeMille}},\ }\href@noop {} {\bibfield  {journal} {\bibinfo  {journal} {Phys.
  Rev. Lett.}\ }\textbf {\bibinfo {volume} {88}},\ \bibinfo {pages} {067901}
  (\bibinfo {year} {2002})}\BibitemShut {NoStop}%
\bibitem [{\citenamefont {Wall}\ and\ \citenamefont {Carr}(2010)}]{Wall:2010}%
  \BibitemOpen
  \bibfield  {author} {\bibinfo {author} {\bibfnamefont {M.~L.}\ \bibnamefont
  {Wall}}\ and\ \bibinfo {author} {\bibfnamefont {L.~D.}\ \bibnamefont
  {Carr}},\ }\href@noop {} {\bibfield  {journal} {\bibinfo  {journal} {Phys.
  Rev. A}\ }\textbf {\bibinfo {volume} {82}},\ \bibinfo {pages} {013611}
  (\bibinfo {year} {2010})}\BibitemShut {NoStop}%
\bibitem [{\citenamefont {Micheli}\ \emph {et~al.}(2006)\citenamefont
  {Micheli}, \citenamefont {Brennen},\ and\ \citenamefont
  {Zoller}}]{Micheli:2006}%
  \BibitemOpen
  \bibfield  {author} {\bibinfo {author} {\bibfnamefont {A.}~\bibnamefont
  {Micheli}}, \bibinfo {author} {\bibfnamefont {G.~K.}\ \bibnamefont
  {Brennen}}, \ and\ \bibinfo {author} {\bibfnamefont {P.}~\bibnamefont
  {Zoller}},\ }\href@noop {} {\bibfield  {journal} {\bibinfo  {journal} {Nature
  Phys.}\ }\textbf {\bibinfo {volume} {2}},\ \bibinfo {pages} {341} (\bibinfo
  {year} {2006})}\BibitemShut {NoStop}%
\bibitem [{\citenamefont {Baron}\ \emph {et~al.}(2012)\citenamefont {Baron},
  \citenamefont {Campbell}, \citenamefont {DeMille}, \citenamefont {Doyle},
  \citenamefont {Gabrielse}, \citenamefont {Gurevich}, \citenamefont {Hess},
  \citenamefont {Hutzler}, \citenamefont {Kirilov}, \citenamefont {Kozyryev},
  \citenamefont {O'Leary}, \citenamefont {Panda}, \citenamefont {Parsons},
  \citenamefont {Petrik}, \citenamefont {Spaun}, \citenamefont {Vutha},\ and\
  \citenamefont {West}}]{Baranov:2012}%
  \BibitemOpen
  \bibfield  {author} {\bibinfo {author} {\bibfnamefont {J.}~\bibnamefont
  {Baron}}, \bibinfo {author} {\bibfnamefont {W.~C.}\ \bibnamefont {Campbell}},
  \bibinfo {author} {\bibfnamefont {D.}~\bibnamefont {DeMille}}, \bibinfo
  {author} {\bibfnamefont {J.~M.}\ \bibnamefont {Doyle}}, \bibinfo {author}
  {\bibfnamefont {G.}~\bibnamefont {Gabrielse}}, \bibinfo {author}
  {\bibfnamefont {Y.~V.}\ \bibnamefont {Gurevich}}, \bibinfo {author}
  {\bibfnamefont {P.~W.}\ \bibnamefont {Hess}}, \bibinfo {author}
  {\bibfnamefont {N.~R.}\ \bibnamefont {Hutzler}}, \bibinfo {author}
  {\bibfnamefont {E.}~\bibnamefont {Kirilov}}, \bibinfo {author} {\bibfnamefont
  {I.}~\bibnamefont {Kozyryev}}, \bibinfo {author} {\bibfnamefont {B.~R.}\
  \bibnamefont {O'Leary}}, \bibinfo {author} {\bibfnamefont {C.~D.}\
  \bibnamefont {Panda}}, \bibinfo {author} {\bibfnamefont {M.~F.}\ \bibnamefont
  {Parsons}}, \bibinfo {author} {\bibfnamefont {E.~S.}\ \bibnamefont {Petrik}},
  \bibinfo {author} {\bibfnamefont {B.}~\bibnamefont {Spaun}}, \bibinfo
  {author} {\bibfnamefont {A.~C.}\ \bibnamefont {Vutha}}, \ and\ \bibinfo
  {author} {\bibfnamefont {A.~D.}\ \bibnamefont {West}},\ }\href@noop {}
  {\bibfield  {journal} {\bibinfo  {journal} {Chem. Rev.}\ }\textbf {\bibinfo
  {volume} {112}},\ \bibinfo {pages} {5012} (\bibinfo {year}
  {2012})}\BibitemShut {NoStop}%
\bibitem [{\citenamefont {Ni}\ \emph {et~al.}(2008)\citenamefont {Ni},
  \citenamefont {Ospelkaus}, \citenamefont {{de Miranda}}, \citenamefont
  {Pe'er}, \citenamefont {Neyenhuis}, \citenamefont {Zirbel}, \citenamefont
  {Kotochigova}, \citenamefont {Julienne}, \citenamefont {Jin},\ and\
  \citenamefont {Ye}}]{Ni:KRb:2008}%
  \BibitemOpen
  \bibfield  {author} {\bibinfo {author} {\bibfnamefont {K.-K.}\ \bibnamefont
  {Ni}}, \bibinfo {author} {\bibfnamefont {S.}~\bibnamefont {Ospelkaus}},
  \bibinfo {author} {\bibfnamefont {M.~H.~G.}\ \bibnamefont {{de Miranda}}},
  \bibinfo {author} {\bibfnamefont {A.}~\bibnamefont {Pe'er}}, \bibinfo
  {author} {\bibfnamefont {B.}~\bibnamefont {Neyenhuis}}, \bibinfo {author}
  {\bibfnamefont {J.~J.}\ \bibnamefont {Zirbel}}, \bibinfo {author}
  {\bibfnamefont {S.}~\bibnamefont {Kotochigova}}, \bibinfo {author}
  {\bibfnamefont {P.~S.}\ \bibnamefont {Julienne}}, \bibinfo {author}
  {\bibfnamefont {D.~S.}\ \bibnamefont {Jin}}, \ and\ \bibinfo {author}
  {\bibfnamefont {J.}~\bibnamefont {Ye}},\ }\href@noop {} {\bibfield  {journal}
  {\bibinfo  {journal} {Science}\ }\textbf {\bibinfo {volume} {322}},\ \bibinfo
  {pages} {231} (\bibinfo {year} {2008})}\BibitemShut {NoStop}%
\bibitem [{\citenamefont {Takekoshi}\ \emph {et~al.}(2014)\citenamefont
  {Takekoshi}, \citenamefont {Reichs\"ollner}, \citenamefont {Schindewolf},
  \citenamefont {Hutson}, \citenamefont {{Le Sueur}}, \citenamefont {Dulieu},
  \citenamefont {Ferlaino}, \citenamefont {Grimm},\ and\ \citenamefont
  {N\"agerl}}]{Takekoshi:RbCs:2014}%
  \BibitemOpen
  \bibfield  {author} {\bibinfo {author} {\bibfnamefont {T.}~\bibnamefont
  {Takekoshi}}, \bibinfo {author} {\bibfnamefont {L.}~\bibnamefont
  {Reichs\"ollner}}, \bibinfo {author} {\bibfnamefont {A.}~\bibnamefont
  {Schindewolf}}, \bibinfo {author} {\bibfnamefont {J.~M.}\ \bibnamefont
  {Hutson}}, \bibinfo {author} {\bibfnamefont {C.~R.}\ \bibnamefont {{Le
  Sueur}}}, \bibinfo {author} {\bibfnamefont {O.}~\bibnamefont {Dulieu}},
  \bibinfo {author} {\bibfnamefont {F.}~\bibnamefont {Ferlaino}}, \bibinfo
  {author} {\bibfnamefont {R.}~\bibnamefont {Grimm}}, \ and\ \bibinfo {author}
  {\bibfnamefont {H.-C.}\ \bibnamefont {N\"agerl}},\ }\href@noop {} {\bibfield
  {journal} {\bibinfo  {journal} {Phys. Rev. Lett.}\ }\textbf {\bibinfo
  {volume} {113}},\ \bibinfo {pages} {205301} (\bibinfo {year}
  {2014})}\BibitemShut {NoStop}%
\bibitem [{\citenamefont {Molony}\ \emph {et~al.}(2014)\citenamefont {Molony},
  \citenamefont {Gregory}, \citenamefont {Ji}, \citenamefont {Lu},
  \citenamefont {K\"oppinger}, \citenamefont {{Le Sueur}}, \citenamefont
  {Blackley}, \citenamefont {Hutson},\ and\ \citenamefont
  {Cornish}}]{Molony:RbCs:2014}%
  \BibitemOpen
  \bibfield  {author} {\bibinfo {author} {\bibfnamefont {P.~K.}\ \bibnamefont
  {Molony}}, \bibinfo {author} {\bibfnamefont {P.~D.}\ \bibnamefont {Gregory}},
  \bibinfo {author} {\bibfnamefont {Z.}~\bibnamefont {Ji}}, \bibinfo {author}
  {\bibfnamefont {B.}~\bibnamefont {Lu}}, \bibinfo {author} {\bibfnamefont
  {M.~P.}\ \bibnamefont {K\"oppinger}}, \bibinfo {author} {\bibfnamefont
  {C.~R.}\ \bibnamefont {{Le Sueur}}}, \bibinfo {author} {\bibfnamefont
  {C.~L.}\ \bibnamefont {Blackley}}, \bibinfo {author} {\bibfnamefont {J.~M.}\
  \bibnamefont {Hutson}}, \ and\ \bibinfo {author} {\bibfnamefont {S.~L.}\
  \bibnamefont {Cornish}},\ }\href@noop {} {\bibfield  {journal} {\bibinfo
  {journal} {Phys. Rev. Lett.}\ }\textbf {\bibinfo {volume} {113}},\ \bibinfo
  {pages} {255301} (\bibinfo {year} {2014})}\BibitemShut {NoStop}%
\bibitem [{\citenamefont {Park}\ \emph {et~al.}(2015)\citenamefont {Park},
  \citenamefont {Will},\ and\ \citenamefont {Zwierlein}}]{Park:2015}%
  \BibitemOpen
  \bibfield  {author} {\bibinfo {author} {\bibfnamefont {J.~W.}\ \bibnamefont
  {Park}}, \bibinfo {author} {\bibfnamefont {S.~A.}\ \bibnamefont {Will}}, \
  and\ \bibinfo {author} {\bibfnamefont {M.~W.}\ \bibnamefont {Zwierlein}},\
  }\href@noop {} {\bibfield  {journal} {\bibinfo  {journal} {Phys. Rev. Lett.}\
  }\textbf {\bibinfo {volume} {114}},\ \bibinfo {pages} {205302} (\bibinfo
  {year} {2015})}\BibitemShut {NoStop}%
\bibitem [{\citenamefont {Guo}\ \emph {et~al.}(2016)\citenamefont {Guo},
  \citenamefont {Zhu}, \citenamefont {Lu}, \citenamefont {Ye}, \citenamefont
  {Wang}, \citenamefont {Vexiau}, \citenamefont {Bouloufa-Maafa}, \citenamefont
  {Qu\'em\'ener}, \citenamefont {Dulieu},\ and\ \citenamefont
  {Wang}}]{Guo:NaRb:2016}%
  \BibitemOpen
  \bibfield  {author} {\bibinfo {author} {\bibfnamefont {M.}~\bibnamefont
  {Guo}}, \bibinfo {author} {\bibfnamefont {B.}~\bibnamefont {Zhu}}, \bibinfo
  {author} {\bibfnamefont {B.}~\bibnamefont {Lu}}, \bibinfo {author}
  {\bibfnamefont {X.}~\bibnamefont {Ye}}, \bibinfo {author} {\bibfnamefont
  {F.}~\bibnamefont {Wang}}, \bibinfo {author} {\bibfnamefont {R.}~\bibnamefont
  {Vexiau}}, \bibinfo {author} {\bibfnamefont {N.}~\bibnamefont
  {Bouloufa-Maafa}}, \bibinfo {author} {\bibfnamefont {G.}~\bibnamefont
  {Qu\'em\'ener}}, \bibinfo {author} {\bibfnamefont {O.}~\bibnamefont
  {Dulieu}}, \ and\ \bibinfo {author} {\bibfnamefont {D.}~\bibnamefont
  {Wang}},\ }\href@noop {} {\bibfield  {journal} {\bibinfo  {journal} {Phys.
  Rev. Lett.}\ }\textbf {\bibinfo {volume} {116}},\ \bibinfo {pages} {205303}
  (\bibinfo {year} {2016})}\BibitemShut {NoStop}%
\bibitem [{\citenamefont {de~Miranda}\ \emph {et~al.}(2011)\citenamefont
  {de~Miranda}, \citenamefont {Chotia}, \citenamefont {Neyenhuis},
  \citenamefont {Wang}, \citenamefont {Qu\'em\'ener}, \citenamefont
  {Ospelkaus}, \citenamefont {Bohn}, \citenamefont {Ye},\ and\ \citenamefont
  {Jin}}]{deMiranda:2011}%
  \BibitemOpen
  \bibfield  {author} {\bibinfo {author} {\bibfnamefont {M.~H.~G.}\
  \bibnamefont {de~Miranda}}, \bibinfo {author} {\bibfnamefont
  {A.}~\bibnamefont {Chotia}}, \bibinfo {author} {\bibfnamefont
  {B.}~\bibnamefont {Neyenhuis}}, \bibinfo {author} {\bibfnamefont
  {D.}~\bibnamefont {Wang}}, \bibinfo {author} {\bibfnamefont {G.}~\bibnamefont
  {Qu\'em\'ener}}, \bibinfo {author} {\bibfnamefont {S.}~\bibnamefont
  {Ospelkaus}}, \bibinfo {author} {\bibfnamefont {J.~L.}\ \bibnamefont {Bohn}},
  \bibinfo {author} {\bibfnamefont {J.}~\bibnamefont {Ye}}, \ and\ \bibinfo
  {author} {\bibfnamefont {D.~S.}\ \bibnamefont {Jin}},\ }\href@noop {}
  {\bibfield  {journal} {\bibinfo  {journal} {Nat. Phys.}\ }\textbf {\bibinfo
  {volume} {7}},\ \bibinfo {pages} {502} (\bibinfo {year} {2011})}\BibitemShut
  {NoStop}%
\bibitem [{\citenamefont {Chotia}\ \emph {et~al.}(2012)\citenamefont {Chotia},
  \citenamefont {Neyenhuis}, \citenamefont {Moses}, \citenamefont {Yan},
  \citenamefont {Covey}, \citenamefont {Foss-Feig}, \citenamefont {Rey},
  \citenamefont {Jin},\ and\ \citenamefont {Ye}}]{Chotia:2012}%
  \BibitemOpen
  \bibfield  {author} {\bibinfo {author} {\bibfnamefont {A.}~\bibnamefont
  {Chotia}}, \bibinfo {author} {\bibfnamefont {B.}~\bibnamefont {Neyenhuis}},
  \bibinfo {author} {\bibfnamefont {S.~A.}\ \bibnamefont {Moses}}, \bibinfo
  {author} {\bibfnamefont {B.}~\bibnamefont {Yan}}, \bibinfo {author}
  {\bibfnamefont {J.~P.}\ \bibnamefont {Covey}}, \bibinfo {author}
  {\bibfnamefont {M.}~\bibnamefont {Foss-Feig}}, \bibinfo {author}
  {\bibfnamefont {A.~M.}\ \bibnamefont {Rey}}, \bibinfo {author} {\bibfnamefont
  {D.~S.}\ \bibnamefont {Jin}}, \ and\ \bibinfo {author} {\bibfnamefont
  {J.}~\bibnamefont {Ye}},\ }\href@noop {} {\bibfield  {journal} {\bibinfo
  {journal} {Phys. Rev. Lett.}\ }\textbf {\bibinfo {volume} {108}},\ \bibinfo
  {pages} {080405} (\bibinfo {year} {2012})}\BibitemShut {NoStop}%
\bibitem [{\citenamefont {Ospelkaus}\ \emph {et~al.}(2010)\citenamefont
  {Ospelkaus}, \citenamefont {Ni}, \citenamefont {Wang}, \citenamefont {{de
  Miranda}}, \citenamefont {Neyenhuis}, \citenamefont {Qu\'{e}m\'{e}ner},
  \citenamefont {Julienne}, \citenamefont {Bohn}, \citenamefont {Jin},\ and\
  \citenamefont {Ye}}]{Ospelkaus:react:2010}%
  \BibitemOpen
  \bibfield  {author} {\bibinfo {author} {\bibfnamefont {S.}~\bibnamefont
  {Ospelkaus}}, \bibinfo {author} {\bibfnamefont {K.-K.}\ \bibnamefont {Ni}},
  \bibinfo {author} {\bibfnamefont {D.}~\bibnamefont {Wang}}, \bibinfo {author}
  {\bibfnamefont {M.~H.~G.}\ \bibnamefont {{de Miranda}}}, \bibinfo {author}
  {\bibfnamefont {B.}~\bibnamefont {Neyenhuis}}, \bibinfo {author}
  {\bibfnamefont {G.}~\bibnamefont {Qu\'{e}m\'{e}ner}}, \bibinfo {author}
  {\bibfnamefont {P.~S.}\ \bibnamefont {Julienne}}, \bibinfo {author}
  {\bibfnamefont {J.~L.}\ \bibnamefont {Bohn}}, \bibinfo {author}
  {\bibfnamefont {D.~S.}\ \bibnamefont {Jin}}, \ and\ \bibinfo {author}
  {\bibfnamefont {J.}~\bibnamefont {Ye}},\ }\href@noop {} {\bibfield  {journal}
  {\bibinfo  {journal} {Science}\ }\textbf {\bibinfo {volume} {327}},\ \bibinfo
  {pages} {853} (\bibinfo {year} {2010})}\BibitemShut {NoStop}%
\bibitem [{\citenamefont {Tscherbul}\ \emph {et~al.}(2010)\citenamefont
  {Tscherbul}, \citenamefont {Calarco}, \citenamefont {Lesanovsky},
  \citenamefont {Krems}, \citenamefont {Dalgarno},\ and\ \citenamefont
  {Schmiedmayer}}]{TVTscherbul:rf:2010}%
  \BibitemOpen
  \bibfield  {author} {\bibinfo {author} {\bibfnamefont {T.~V.}\ \bibnamefont
  {Tscherbul}}, \bibinfo {author} {\bibfnamefont {T.}~\bibnamefont {Calarco}},
  \bibinfo {author} {\bibfnamefont {I.}~\bibnamefont {Lesanovsky}}, \bibinfo
  {author} {\bibfnamefont {R.~V.}\ \bibnamefont {Krems}}, \bibinfo {author}
  {\bibfnamefont {A.}~\bibnamefont {Dalgarno}}, \ and\ \bibinfo {author}
  {\bibfnamefont {J.}~\bibnamefont {Schmiedmayer}},\ }\href {\doibase
  10.1103/PhysRevA.81.050701} {\bibfield  {journal} {\bibinfo  {journal} {Phys.
  Rev. A}\ }\textbf {\bibinfo {volume} {81}},\ \bibinfo {pages} {050701(R)}
  (\bibinfo {year} {2010})}\BibitemShut {NoStop}%
\bibitem [{\citenamefont {Hanna}\ \emph {et~al.}(2010)\citenamefont {Hanna},
  \citenamefont {Tiesinga},\ and\ \citenamefont {Julienne}}]{Hanna:2010}%
  \BibitemOpen
  \bibfield  {author} {\bibinfo {author} {\bibfnamefont {T.~M.}\ \bibnamefont
  {Hanna}}, \bibinfo {author} {\bibfnamefont {E.}~\bibnamefont {Tiesinga}}, \
  and\ \bibinfo {author} {\bibfnamefont {P.~S.}\ \bibnamefont {Julienne}},\
  }\href@noop {} {\bibfield  {journal} {\bibinfo  {journal} {New J. Phys.}\
  }\textbf {\bibinfo {volume} {12}},\ \bibinfo {pages} {083031} (\bibinfo
  {year} {2010})}\BibitemShut {NoStop}%
\bibitem [{\citenamefont {Smith}(2015)}]{Smith:rf:2015}%
  \BibitemOpen
  \bibfield  {author} {\bibinfo {author} {\bibfnamefont {D.~H.}\ \bibnamefont
  {Smith}},\ }\href@noop {} {\bibfield  {journal} {\bibinfo  {journal} {Phys.
  Rev. Lett.}\ }\textbf {\bibinfo {volume} {115}},\ \bibinfo {pages} {193002}
  (\bibinfo {year} {2015})}\BibitemShut {NoStop}%
\bibitem [{\citenamefont {Thompson}\ \emph {et~al.}(2005)\citenamefont
  {Thompson}, \citenamefont {Hodby},\ and\ \citenamefont
  {Wieman}}]{Thompson:magres:2005}%
  \BibitemOpen
  \bibfield  {author} {\bibinfo {author} {\bibfnamefont {S.~T.}\ \bibnamefont
  {Thompson}}, \bibinfo {author} {\bibfnamefont {E.}~\bibnamefont {Hodby}}, \
  and\ \bibinfo {author} {\bibfnamefont {C.~E.}\ \bibnamefont {Wieman}},\
  }\href@noop {} {\bibfield  {journal} {\bibinfo  {journal} {Phys. Rev. Lett.}\
  }\textbf {\bibinfo {volume} {95}},\ \bibinfo {pages} {190404} (\bibinfo
  {year} {2005})}\BibitemShut {NoStop}%
\bibitem [{\citenamefont {Zirbel}\ \emph {et~al.}(2008)\citenamefont {Zirbel},
  \citenamefont {Ni}, \citenamefont {Ospelkaus}, \citenamefont {Nicholson},
  \citenamefont {Olsen}, \citenamefont {Julienne}, \citenamefont {Wieman},
  \citenamefont {Ye},\ and\ \citenamefont {Jin}}]{Zirbel:2008}%
  \BibitemOpen
  \bibfield  {author} {\bibinfo {author} {\bibfnamefont {J.~J.}\ \bibnamefont
  {Zirbel}}, \bibinfo {author} {\bibfnamefont {K.-K.}\ \bibnamefont {Ni}},
  \bibinfo {author} {\bibfnamefont {S.}~\bibnamefont {Ospelkaus}}, \bibinfo
  {author} {\bibfnamefont {T.~L.}\ \bibnamefont {Nicholson}}, \bibinfo {author}
  {\bibfnamefont {M.~L.}\ \bibnamefont {Olsen}}, \bibinfo {author}
  {\bibfnamefont {P.~S.}\ \bibnamefont {Julienne}}, \bibinfo {author}
  {\bibfnamefont {C.~E.}\ \bibnamefont {Wieman}}, \bibinfo {author}
  {\bibfnamefont {J.}~\bibnamefont {Ye}}, \ and\ \bibinfo {author}
  {\bibfnamefont {D.~S.}\ \bibnamefont {Jin}},\ }\href@noop {} {\bibfield
  {journal} {\bibinfo  {journal} {Phys. Rev. A}\ }\textbf {\bibinfo {volume}
  {78}},\ \bibinfo {pages} {013416} (\bibinfo {year} {2008})}\BibitemShut
  {NoStop}%
\bibitem [{\citenamefont {Patel}\ \emph {et~al.}(2014)\citenamefont {Patel},
  \citenamefont {Blackley}, \citenamefont {Cornish},\ and\ \citenamefont
  {Hutson}}]{Patel:2014}%
  \BibitemOpen
  \bibfield  {author} {\bibinfo {author} {\bibfnamefont {H.~J.}\ \bibnamefont
  {Patel}}, \bibinfo {author} {\bibfnamefont {C.~L.}\ \bibnamefont {Blackley}},
  \bibinfo {author} {\bibfnamefont {S.~L.}\ \bibnamefont {Cornish}}, \ and\
  \bibinfo {author} {\bibfnamefont {J.~M.}\ \bibnamefont {Hutson}},\
  }\href@noop {} {\bibfield  {journal} {\bibinfo  {journal} {Phys. Rev. A}\
  }\textbf {\bibinfo {volume} {90}},\ \bibinfo {pages} {032716} (\bibinfo
  {year} {2014})}\BibitemShut {NoStop}%
\bibitem [{\citenamefont {Ferber}\ \emph {et~al.}(2013)\citenamefont {Ferber},
  \citenamefont {Nikolayeva}, \citenamefont {Tamanis}, \citenamefont
  {Kn\"ockel},\ and\ \citenamefont {Tiemann}}]{Ferber:2013}%
  \BibitemOpen
  \bibfield  {author} {\bibinfo {author} {\bibfnamefont {R.}~\bibnamefont
  {Ferber}}, \bibinfo {author} {\bibfnamefont {O.}~\bibnamefont {Nikolayeva}},
  \bibinfo {author} {\bibfnamefont {M.}~\bibnamefont {Tamanis}}, \bibinfo
  {author} {\bibfnamefont {H.}~\bibnamefont {Kn\"ockel}}, \ and\ \bibinfo
  {author} {\bibfnamefont {E.}~\bibnamefont {Tiemann}},\ }\href@noop {}
  {\bibfield  {journal} {\bibinfo  {journal} {Phys. Rev. A}\ }\textbf {\bibinfo
  {volume} {88}},\ \bibinfo {pages} {012516} (\bibinfo {year}
  {2013})}\BibitemShut {NoStop}%
\bibitem [{\citenamefont {Berninger}\ \emph {et~al.}(2013)\citenamefont
  {Berninger}, \citenamefont {Zenesini}, \citenamefont {Huang}, \citenamefont
  {Harm}, \citenamefont {N\"agerl}, \citenamefont {Ferlaino}, \citenamefont
  {Grimm}, \citenamefont {Julienne},\ and\ \citenamefont
  {Hutson}}]{Berninger:Cs2:2013}%
  \BibitemOpen
  \bibfield  {author} {\bibinfo {author} {\bibfnamefont {M.}~\bibnamefont
  {Berninger}}, \bibinfo {author} {\bibfnamefont {A.}~\bibnamefont {Zenesini}},
  \bibinfo {author} {\bibfnamefont {B.}~\bibnamefont {Huang}}, \bibinfo
  {author} {\bibfnamefont {W.}~\bibnamefont {Harm}}, \bibinfo {author}
  {\bibfnamefont {H.-C.}\ \bibnamefont {N\"agerl}}, \bibinfo {author}
  {\bibfnamefont {F.}~\bibnamefont {Ferlaino}}, \bibinfo {author}
  {\bibfnamefont {R.}~\bibnamefont {Grimm}}, \bibinfo {author} {\bibfnamefont
  {P.~S.}\ \bibnamefont {Julienne}}, \ and\ \bibinfo {author} {\bibfnamefont
  {J.~M.}\ \bibnamefont {Hutson}},\ }\href@noop {} {\bibfield  {journal}
  {\bibinfo  {journal} {Phys. Rev. A}\ }\textbf {\bibinfo {volume} {87}},\
  \bibinfo {pages} {032517} (\bibinfo {year} {2013})}\BibitemShut {NoStop}%
\bibitem [{\citenamefont {Hutson}\ and\ \citenamefont
  {Green}(1994)}]{molscat:v14}%
  \BibitemOpen
  \bibfield  {author} {\bibinfo {author} {\bibfnamefont {J.~M.}\ \bibnamefont
  {Hutson}}\ and\ \bibinfo {author} {\bibfnamefont {S.}~\bibnamefont {Green}},\
  }\href@noop {} {\enquote {\bibinfo {title} {{MOLSCAT} computer program,
  version 14},}\ }\bibinfo {howpublished} {distributed by Collaborative
  Computational Project No.\ 6 of the UK Engineering and Physical Sciences
  Research Council} (\bibinfo {year} {1994})\BibitemShut {NoStop}%
\bibitem [{\citenamefont {Hutson}(1993)}]{Hutson:bound:1993}%
  \BibitemOpen
  \bibfield  {author} {\bibinfo {author} {\bibfnamefont {J.~M.}\ \bibnamefont
  {Hutson}},\ }\href@noop {} {\enquote {\bibinfo {title} {{BOUND} computer
  program, version 5},}\ }\bibinfo {howpublished} {distributed by Collaborative
  Computational Project No.\ 6 of the UK Engineering and Physical Sciences
  Research Council} (\bibinfo {year} {1993})\BibitemShut {NoStop}%
\bibitem [{\citenamefont {Hutson}(2011)}]{Hutson:field:2011}%
  \BibitemOpen
  \bibfield  {author} {\bibinfo {author} {\bibfnamefont {J.~M.}\ \bibnamefont
  {Hutson}},\ }\href@noop {} {\enquote {\bibinfo {title} {{FIELD} computer
  program, version 1},}\ } (\bibinfo {year} {2011})\BibitemShut {NoStop}%
\bibitem [{\citenamefont {Hutson}\ \emph {et~al.}(2008)\citenamefont {Hutson},
  \citenamefont {Tiesinga},\ and\ \citenamefont
  {Julienne}}]{Hutson:Cs2-note:2008}%
  \BibitemOpen
  \bibfield  {author} {\bibinfo {author} {\bibfnamefont {J.~M.}\ \bibnamefont
  {Hutson}}, \bibinfo {author} {\bibfnamefont {E.}~\bibnamefont {Tiesinga}}, \
  and\ \bibinfo {author} {\bibfnamefont {P.~S.}\ \bibnamefont {Julienne}},\
  }\href@noop {} {\bibfield  {journal} {\bibinfo  {journal} {Phys. Rev. A}\
  }\textbf {\bibinfo {volume} {78}},\ \bibinfo {pages} {052703} (\bibinfo
  {year} {2008})},\ \bibinfo {note} {note that the matrix element of the
  dipolar spin-spin operator given in Eq.\ A2 of this paper omits a factor of
  $-\sqrt{30}$.}\BibitemShut {Stop}%
\bibitem [{\citenamefont {Arimondo}\ \emph {et~al.}(1977)\citenamefont
  {Arimondo}, \citenamefont {Inguscio},\ and\ \citenamefont
  {Violino}}]{Arimondo:1977}%
  \BibitemOpen
  \bibfield  {author} {\bibinfo {author} {\bibfnamefont {E.}~\bibnamefont
  {Arimondo}}, \bibinfo {author} {\bibfnamefont {M.}~\bibnamefont {Inguscio}},
  \ and\ \bibinfo {author} {\bibfnamefont {P.}~\bibnamefont {Violino}},\
  }\href@noop {} {\bibfield  {journal} {\bibinfo  {journal} {Rev. Mod. Phys.}\
  }\textbf {\bibinfo {volume} {49}},\ \bibinfo {pages} {31} (\bibinfo {year}
  {1977})}\BibitemShut {NoStop}%
\bibitem [{\citenamefont {Mordovin}(2015)}]{Mordovin:2015}%
  \BibitemOpen
  \bibfield  {author} {\bibinfo {author} {\bibfnamefont {I.}~\bibnamefont
  {Mordovin}},\ }\emph {\bibinfo {title} {Radio-Frequency Induced Association
  of Molecules in $^{87}$Rb}},\ \href@noop {} {Ph.D. thesis},\ \bibinfo
  {school} {Swinburne University of Technology}, \bibinfo {address} {Melbourne}
  (\bibinfo {year} {2015})\BibitemShut {NoStop}%
\bibitem [{\citenamefont {Morizot}\ \emph {et~al.}(2008)\citenamefont
  {Morizot}, \citenamefont {Longchambon}, \citenamefont {Kollengode~Easwaran},
  \citenamefont {Dubessy}, \citenamefont {Knyazchyan}, \citenamefont {Pottie},
  \citenamefont {Lorent},\ and\ \citenamefont {Perrin}}]{Morizot:2008}%
  \BibitemOpen
  \bibfield  {author} {\bibinfo {author} {\bibfnamefont {O.}~\bibnamefont
  {Morizot}}, \bibinfo {author} {\bibfnamefont {L.}~\bibnamefont
  {Longchambon}}, \bibinfo {author} {\bibfnamefont {R.}~\bibnamefont
  {Kollengode~Easwaran}}, \bibinfo {author} {\bibfnamefont {R.}~\bibnamefont
  {Dubessy}}, \bibinfo {author} {\bibfnamefont {E.}~\bibnamefont {Knyazchyan}},
  \bibinfo {author} {\bibfnamefont {P.-E.}\ \bibnamefont {Pottie}}, \bibinfo
  {author} {\bibfnamefont {V.}~\bibnamefont {Lorent}}, \ and\ \bibinfo {author}
  {\bibfnamefont {H.}~\bibnamefont {Perrin}},\ }\href@noop {} {\bibfield
  {journal} {\bibinfo  {journal} {Eur. Phys. J. D}\ }\textbf {\bibinfo {volume}
  {47}},\ \bibinfo {pages} {209} (\bibinfo {year} {2008})}\BibitemShut
  {NoStop}%
\bibitem [{\citenamefont {Hutson}(2007)}]{Hutson:res:2007}%
  \BibitemOpen
  \bibfield  {author} {\bibinfo {author} {\bibfnamefont {J.~M.}\ \bibnamefont
  {Hutson}},\ }\href@noop {} {\bibfield  {journal} {\bibinfo  {journal} {New J.
  Phys.}\ }\textbf {\bibinfo {volume} {9}},\ \bibinfo {pages} {152} (\bibinfo
  {year} {2007})}\BibitemShut {NoStop}%
\end{thebibliography}%
\end{document}